\documentclass[final,3p,times,twocolumn]{elsarticle}

\biboptions{sort&compress}
\usepackage{amssymb,mathrsfs}
\usepackage{amsthm,amsmath}
\usepackage{epstopdf}
\usepackage{bm}% bold math
\usepackage{multirow}
\usepackage{slashed}
\usepackage{threeparttable}

\usepackage[%dvipdfm,
bookmarks=true,colorlinks,%
citecolor=blue,linkcolor=blue,%hypertex, %
breaklinks=true]{hyperref}

\begin{document}
	
	%\begin{CJK*}{GBK}{song}
	
	\hyphenpenalty=6000
	\tolerance=3000
	
	\begin{frontmatter}
		
		%% Title, authors and addresses
		
		%% use the tnoteref command within \title for footnotes;
		%% use the tnotetext command for theassociated footnote;
		%% use the fnref command within \author or \address for footnotes;
		%% use the fntext command for theassociated footnote;
		%% use the corref command within \author for corresponding author footnotes;
		%% use the cortext command for theassociated footnote;
		%% use the ead command for the email address,
		%% and the form \ead[url] for the home page:
		%% \title{Title\tnoteref{label1}}
		%% \tnotetext[label1]{}
		%% \author{Name\corref{cor1}\fnref{label2}}
		%% \ead{email address}
		%% \ead[url]{home page}
		%% \fntext[label2]{}
		%% \cortext[cor1]{}
		
		%\title{Octupole correlations in $^{96}$Zr from angular momentum and parity projections}
        %\title{Anatomy of $^{96}$Zr potential energy surfaces with octupole correlations from angular momentum and parity projections}
        \title{Anatomy of octupole correlations in $^{96}$Zr with a symmetry-restored multidimensionally-constrained covariant density functional theory}
		
		%% use optional labels to link authors explicitly to addresses:
		
		\author[GXNU,ITP] {Yu-Ting Rong}
		%{\color{red}
		\author[JXNU] {Xian-Ye Wu}
	%}
		\author[CAEP]{Bing-Nan Lu \corref{cor1}}
		\ead{bnlv@gscaep.ac.cn}
		\cortext[cor1]{Corresponding author.}
		\author[SYSU] {Jiang-Ming Yao}
		
		\affiliation[GXNU]{%
			organization={Guangxi Key Laboratory of Nuclear Physics and Technology, Guangxi Normal University},%
			city={Guilin},%
			postcode={541004},%
			country={China}}
		
		\affiliation[ITP]{%
			organization={CAS Key Laboratory of Theoretical Physics, Institute of Theoretical Physics,
				Chinese Academy of Sciences},%
			%addressline={P.O. Box 2735},%
			city={Beijing},%
			postcode={100190},%
			%state={},%
			country={China}}
		
		\affiliation[JXNU]{%
			organization={College of Physics and Communication Electronics, Jiangxi Normal University},%
			city={Nanchang},%
			postcode={330022},%
			country={China}}
		
		\affiliation[CAEP]{%
			organization={Graduate School of China Academy of Engineering Physics},%
			city={Beijing},%
			postcode={100193},%
			country={China}}
		
		\affiliation[SYSU]{%
			organization={School of Physics and Astronomy, Sun Yat-sen University},%
			city={Zhuhai},%
			postcode={519082},%
			country={China}}

		\begin{abstract}
        A recent analysis based on the STAR measurement in relativistic heavy-ion collision experiments provides
        evidence of octupole correlation in the ground state of $^{96}$Zr, the description of which presents a challenge
        to nuclear structure models. In this work, we perform projection-after-variation calculations for $^{96}$Zr
        based on a multidimensionally-constrained relativistic Hartree-Bogoliubov model. Our results show that
        an octupole deformed shape is favored in energy after symmetry restoration, and this phenomenon
        cannot be reproduced in the pure mean-field calculations. These complex structures originate from the
        competition among various shell structures in this mass region. Our results suggest that the allowance
        of symmetry breaking at the mean-field level and the restoration of broken symmetries are essential
        elements for understanding the structure of $^{96}$Zr in density functional theories.
	
		\end{abstract}

		
	\end{frontmatter}
	
\section{INTRODUCTION}\label{sec1}

%八极形变的实验和理论进展
%The nuclear intrinsic shapes can be characterized by a few multipole deformation parameters $\beta_{\lambda \mu}$~\cite{Heyde2011_RMP83-1467,Heyde2016_PS91-083008}.

In atomic nuclei, the occurrence of spontaneous symmetry breaking results in various shapes.
The most common nuclear shape is the ellipsoid.
For some specific proton or neutron numbers, we can also find pronounced pear-like shapes
~\cite{Butler1996_RMP68-349,Butler2020_PRSA476-20200202,Moeller2008_ADNDT94-758,Robledo2011_PRC84-054302,Agbemava2016_PRC93-044304,Cao2020_PRC102-024311}.
For a long time, the only reliable method for studying these reflection-asymmetric shapes has been to search for the characteristic patterns of parity and angular momentum in nuclear spectroscopy.
For example, the enhanced reduced electric-octupole transition probabilities, $B(E3)$, in $^{224}$Ra~\cite{Gaffney2013_Nature497-199}, $^{144}$Ba~\cite{Bucher2016_PRL116-112503}, $^{146}$Ba~\cite{Bucher2017_PRL118-152504}, and $^{228}$Th~\cite{Chishti2020_NP16-853} can be viewed as direct experimental evidences of strong octupole deformations.

Recently, it has been proposed that nuclear deformations can also be extracted from relativistic heavy-ion collisions (RHIC) by analysing the hydrodynamic collective flow of the final-state particles~\cite{Heinz2005_PRL94-132301,Shou2015_PLB749-215,Giacalone2021_PRC104-L041903,Jia2022_PRC105-014905,Giacalone2021_PRL127-242301}. Later this method was applied to study realistic experimental data~\cite{Adamczyk2015_PRL115-222301,Acharya2018_PLB784-82,Sirunyan2019_PRC100-044902,Aad2020_PRC101-024906}.
In Ref.~\cite{Zhang2022_PRL128-022301}, the deformations of $^{96}$Zr and $^{96}$Ru were extracted from isobaric $^{96}$Zr+$^{96}$Zr and $^{96}$Ru+$^{96}$Ru collisions, respectively.
By analysing the elliptic flow $v_2$ and triangular
flow $v_3$~\cite{STAR2021_arXiv2109.00131}, the authors inferred a strong octupole deformation in $^{96}$Zr and a strong quadrupole deformation in $^{96}$Ru.

The onset of large octupole correlation in the low-lying states of atomic nuclei is attributed to the existence of pairs of single-particle orbitals with opposite parities strongly coupled by the octupole moments near the Fermi level~\cite{Butler1996_RMP68-349}.
Consequently, nuclei with neutron or proton numbers near the octupole magic numbers 34, 56, 88, and 134 are natural candidates for searching for the pear-like nucleus.
$^{96}$Zr, whose neutron number is $N=56$, is one of these nuclei and expected to show reflection-asymmetric shapes~\cite{Mach1990_PRC42-R811,Ohm1990_PLB241-472,Hofer1993_NPA551-173,Horen1993_PRC48-R2131,Fayans1994_NPA577-557,Rosso1993_NPA563-74,Skalski1993_NPA559-221,Iskra2019_PLB788-396}.
The experimentally observed large $B(E3)$ value in its ground-state band indicates a strong octupole collectivity~\cite{Mach1990_PRC42-R811,Ohm1990_PLB241-472,Hofer1993_NPA551-173,Horen1993_PRC48-R2131,Iskra2019_PLB788-396},
which is also corroborated by recent RHIC experiments~\cite{Zhang2022_PRL128-022301}.
Conversely, all systematic search for the octupole deformations based on modern nuclear structure theories, including the macroscopic-microscopic (MM) model~\cite{Moeller2016_ADNDT109-1}, the relativistic
Hartree-Bogoliubov (RHB) theory~\cite{Agbemava2016_PRC93-044304}, the Hartree-Fock-Bogoliubov (HFB) theories with Gogny interaction~\cite{Robledo2011_PRC84-054302} and Skyrme interactions~\cite{Cao2020_PRC102-024311}, do not find any octupole deformation for $^{96}$Zr.
Instead, all these theoretical investigations except for the MM model find octupole deformations near $N=Z=40$, which is incompatible with both the experiments and the argument of octupole magic numbers.
To reconcile this inconsistency between the theories and experiments, we need to consider the beyond-mean-field effects not included in the systematic calculations.

The nuclear density functional theories are very successful in describing the nuclear deformations microscopically.
In these calculations, the ground states are obtained by applying the variational principle with a Slater determinant or a Bogoliubov vacuum.
The resulting states usually violate fundamental symmetries and conservation laws, such as rotational symmetry, translational invariance and particle-number conservation \cite{Yao2022_PPNP-103965}.
As a consequence, projection-after-variation (PAV) techniques are employed to restore the broken symmetries and calculate observables with good quantum numbers~\cite{Bender2003_RMP75-121,Niksic2011_PPNP66-519,Egido2016_PS91-073003,Robledo2018_JPG46-013001,Sheikh2021_JPG48-123001,Sun2021_SciBulletin66-2072}.
PAV calculations have been applied to study the octupole deformations in Ra isotopes based on shell model wave functions~\cite{Chen2000_PRC63-014314}, in Ra isotopes~\cite{Yao2015_PRC92-041304} and $^{20}$Ne~\cite{Zhou2016_PLB753-227} based on Dirac wave functions, in Ba, Ra isotopes~\cite{Egido1991_NPA524-65, Bernard2016_PRC93-061302}, Zr isotopes~\cite{Tagami2015_JPG42-015106}, $^{16}$O~\cite{Wang2019_PLB790-498}, actinides,  superheavies~\cite{Rodriguez-Guzman2021_PRC103-044301} and other nuclei~\cite{Robledo2016_EPJA52-300} based on HFB wave functions.
There are also theoretical investigations of octupole deformations based on the multidimensional collective Hamiltonian~\cite{Li2013_PLB726-866,Xia2017_PRC96-054303} and the interacting boson model~\cite{Nomura2021_PRC104-044324}.
The octupole deformations in Zr isotopes have been discussed using the generator-coordinate method (GCM) with a partial angular momentum~\cite{Skalski1993_NPA559-221} or parity~\cite{Zberecki2006_PRC74-051302R,Zberecki2009_PRC79-014319} projection.
Besides the axial deformations, recent PAV calculations based on Gogny interaction predicted a tetrahedral shape associated with a pure non-axial $\beta_{32}$ deformation in $^{96}$Zr~\cite{Tagami2015_JPG42-015106}.
Nevertheless, so far there is no systematic investigations for the evolution of octupole deformations in the $A\sim 100$ region based on full angular momentum and parity projections.
Such calculations will provide necessary knowledge that can help understand the recent RHIC experiments as well as the spectroscopic data.

For studying the various nuclear shapes in a unified framework, the multidimensionally-constrained relativistic Hartree-Bogoliubov (MDCRHB) model has been developed, in which both the axial and spatial reflection symmetries are broken but the simplex symmetry is preserved ~\cite{Lu2014_PRC89-014323,Zhou2016_PS91-063008}. 
Recently we developed a projected multidimensionally-constrained relativistic Hartree-Bogoliubov (p-MDCRHB) model by incorporating the parity and angular momentum projections into the MDCRHB model to study the low-lying states related to exotic nuclear shapes, e.g., the triangular shape associated with the three-$\alpha$ configuration in $^{12}$C~\cite{Wang2022_CTP74-015303}.

In this paper, we investigate the octupole correlations in $^{96}$Zr by employing the p-MDCRHB model. We present a systematic beyond-mean-field study with both triaxial and octupole shapes included simultaneously. In Sec. \ref{sec:model}, the theoretical framework of this model is introduced. In Sec. \ref{sec:results}, we use the p-MDCRHB model to calculate the projected PESs in $^{96}$Zr and extract the deformations and excitation energies of the low-lying states. %Comparison with spectroscopic measurements and relativistic collisions experiments are discussed. 
The results are discussed in comparison with spectroscopic measurements and relativistic collision experiments.
The microscopic mechanism of the octupole correlations are analysed based on the evolution of single-particle shell structures. Finally, a summary is given in Sec. \ref{sec:summary}.

\section{Theoretical framework}\label{sec:model}
%点耦合的Dirac equ
We start with the RHB theory with effective point coupling interactions~\cite{Nikolaus1992_PRC46-1757,Buervenich2002_PRC65-044308}. 
The RHB equation in coordinate space reads~\cite{Ring1996_PPNP37-193,Kucharek1991_ZPA339-23}
\begin{equation}
	\label{eq:rhb}
	\int d^{3}\bm{r}^{\prime}
	\left( \begin{array}{cc} h-\lambda  &  \Delta                      \\
		-\Delta^{*}   & -h+\lambda \end{array}
	\right)
	\left( \begin{array}{c} U_{k} \\ V_{k} \end{array} \right)
	= E_{k}
	\left( \begin{array}{c} U_{k} \\ V_{k} \end{array} \right),
\end{equation}
where $\Delta$ is the pairing field, $\lambda$ is the Fermi energy, $E_k$ is the quasi-particle energy and $(U_k(\bm{r}),V_k(\bm{r}))^T$ is the quasi-particle wave function.
$h$ is the single-particle Hamiltonian,
\begin{equation}
	h = \bm{\alpha} \cdot \bm{p}  +
	\beta \left[ M + S(\bm{r}) \right]+ V(\bm{r}),
\end{equation}
where $M$ is the nucleon mass, $S(\bm{r})$ is the scalar potential, and $V(\bm{r})$ is the vector potential.
%In the MDCRHB model we solve the RHB equation by expanding the wave functions on a axially-deformed harmonic oscillator basis. 
In this work, we use the PC-PK1 parameter set~\cite{Zhao2010_PRC82-054319} and solve the RHB equation by expanding the wave functions on an axially-deformed harmonic oscillator basis.
For pairing interaction, we adopt a separable pairing force of finite range in the spin-singlet channel~\cite{Tian2006_CPL23-3226,Tian2009_PLB676-44},
%\begin{equation}
%\label{eq:separable}
%V(\bm{r}_{1}         \sigma_{1},         \bm{r}_{2}         \sigma_{2},
%\bm{r}_{1}^{\prime}\sigma_{1}^{\prime},\bm{r}_{2}^{\prime}\sigma_{2}^{\prime})
%=
%-G \delta(\bm{R}-\bm{R}^{\prime}) P(\bm{r}) P(\bm{r}^{\prime})
%\frac{1-P_{\sigma}}{2},
%\end{equation}
\begin{equation}
	\label{eq:separable}
	V=-G \delta(\bm{R}-\bm{R}^{\prime}) P(\bm{r}) P(\bm{r}^{\prime})
	\frac{1-P_{\sigma}}{2},
\end{equation}
where $G$ is the pairing strength,
$\bm{R}=(\bm{r}_{1}+\bm{r}_{2}) / 2$ and $\bm{r}=\bm{r}_{1}-\bm{r}_{2}$
are the center of mass and relative coordinates, respectively. The symbols with and without the prime denote the quantities before and after the interaction, respectively.
$P_\sigma$ is the spin-exchange operator.
$P(\bm{r}) = ( 4\pi a^2 )^{-3/2} e^{-{r^2}/{4 a^2}}$ is the Gaussian function.
$a = 0.644$ fm is the effective range of the pairing force.
%\begin{equation}
%P(\bm{r}) = {\left( 4\pi a^2 \right)^{-3/2}} e^{-{r^2}/{4 a^2}},
%\end{equation}
%with $a$ the effective range of the pairing force.
In this work we use different pairing strength for neutron and proton, $G_n=728.00$ MeV$\cdot$fm$^3$ and  $G_p=815.36$ MeV$\cdot$fm$^3$.
%The effective range of the pairing force is fixed to $a=0.644$ fm.
These parameters were adjusted to reproduce the available empirical pairing gaps of $^{102,104}$Zr~\cite{Zhao2017_PRC95-014320}.

%形变约束
%To obtain a potential energy surface (PES), we use a modified linear constraint method~\cite{Lu2014_PRC89-014323} in which the Routhian is defined as
%\begin{equation}
%	E'=E_{\text{RMF}}+\sum_{\lambda\mu}C_{\lambda\mu}Q_{\lambda\mu},
%\end{equation}
%where $E_{\text{RMF}}$ is the mean-field energy, $C_{\lambda\mu}$ is Lagrangian multiplier varied during the iteration and $Q_{\lambda\mu}$ is the multipole moment of the intrinsic densities.
%\begin{equation}
%Q_{\lambda\mu}=\int d^3r\rho_V(\bm{r}) r^\lambda Y_{\lambda\mu}(\Omega),
%\end{equation}
%with $\rho_V(\bm{r})$ the vector density and $Y_{\lambda \mu}(\Omega)$ the spherical harmonics.
Nuclear shapes are characterized by the deformation
parameters $\beta_{\lambda\mu}$ which are defined as
\begin{equation}
	\beta_{\lambda\mu}=\dfrac{4\pi}{3A R^\lambda}Q_{\lambda\mu},
\end{equation}
where $R=1.2A^{1/3}$ fm with $A$ the mass number, $Q_{\lambda\mu}$ is the multipole moment of the intrinsic densities.
With symmetry imposed in this model, all $\beta_{\lambda\mu}$ with even $\mu$ can be considered simultaneously.

%投影波函数的构造，Hill–Wheeler–Griffin equation
%The wave functions with good quantum numbers can be obtained by the p-MDCRHB model~\cite{Wang2021}.
In the p-MDCRHB model~\cite{Wang2022_CTP74-015303}, the nuclear wave function $|\Phi(q)\rangle$ is projected onto the parity $\pi$ and angular momentum $J$ %we make parity and angular momentum projections on a RHB wave function $|\Phi(q)\rangle$,
\begin{equation}
	|\Psi_{\alpha,q}^{JM\pi}\rangle=\sum_K f_\alpha^{JK\pi}\hat{P}^{J}_{MK}\hat{P}^\pi |\Phi(q)\rangle,
\end{equation}
where $f_\alpha^{JK\pi}$ is the weight
function, $q$ represents a collection of the deformation parameters,
\begin{equation}
	\hat{P}_{MK}^J
	=
	\frac{2J + 1}{8 \pi^2} \int {\rm d} \Omega D_{MK}^{J*} (\Omega) \hat{R} (\Omega)
\end{equation}
is {an operator which projects out the component with angular momentum $J$ and its projection $M$ from the deformed mean-field wave function}. $K$ is angular momentum projection onto $z$-axis in the intrinsic frame.
$D_{MK}^J (\Omega)$ is the Wigner function with a Euler angle $\Omega \equiv \{ \phi, \theta, \psi \}$.
%$I$ is the total angular momentum, 
 $\hat{R}$ represents spatial rotation.
The parity projection operator writes
\begin{equation}
	\hat{P}^\pi
	=
	\frac{1}{2} (1 + \pi \hat{P}),
\end{equation}
where $\hat{P}$ is the spatial reflection operation. %and $\pi=\pm 1$ is the parity.

 Considering that the projection calculation with both rotational and parity symmetry breaking is already costly, in this first step towards full projection, we do not include the exact particle number projection. Instead, we add two correction terms to the Hamiltonian kernel as in Ref.~\cite{Yao2010_PRC81-044311} to approximately restore the average proton and neutron numbers, i.e.,
\begin{equation}
{\mathcal H'}={\mathcal H}-\lambda_p[Z(\bm{r};q,q';\Omega)-Z_0]-\lambda_n[N(\bm{r};q,q';\Omega)-N_0],
\label{eq:PNrestoration}
\end{equation}
where $\lambda_p$ and $\lambda_n$ are the Fermi energies for protons and neutrons of the mean-field state $|\Phi(q)\rangle$, $Z(\bm{r};q,q';\Omega)$ and $N(\bm{r};q,q';\Omega)$ are the mixed vector densities for protons and neutrons, respectively.  $Z_0$ and $N_0$ are the target proton and neutron numbers, respectively.
After calculating the mixed energy density, the weight function $f_\alpha^{JK\pi}$ and the eigenvalue $E_\alpha^{J\pi}$ are obtained by solving the generalized eigenvalue equation in the standard way \cite{Ring1980,Yao2009_PRC79-044312}:
\begin{equation}
	\sum_{K'}\left\{{\cal H'}_{KK'}^{J\pi}(q;q)-E_{\alpha}^{J\pi} {\cal N}_{KK'}^{J\pi}(q;q)\right\}f_{\alpha}^{JK'\pi}=0,
\end{equation}
 where $\cal N$ is the norm kernel. 
To check the validity of the ansatz of introducing the correction terms in Eq.~(\ref{eq:PNrestoration}), we also carry out a beyond relativistic mean-field calculation with exact particle-number projections by employing the approach developed in Ref. \cite{Yao2015_PRC92-041304}.

We note that the restoration of broken symmetries with modern nuclear energy density functionals usually meet the problems of spurious divergencies~\cite{Anguiano2001_NPA696-476,Tajima1992_NPA542-355} and finite steps~\cite{Bender2009_PRC79-044319,Duguet2009_PRC79-044320,Dobaczewski2007_PRC76-054315,Lacroix2009_PRC79-044318}. The solution to this problem in a proper way is still an open question. Of course, these issues can be avoided in Hamiltonian-based frameworks if all the terms in the projected energy are taken into account in a consistent way, in particular Fock terms and contributions of the Coulomb and spin-orbit potentials to pairing, etc.~\cite{Anguiano2001_NPA696-476,Doenau1998_PRC58-872,Almehed2001_PRC63-044311}. In the current covariant density functional theories, these problems remain in one way or another. They are not serious in this work as the symmetry restoration is carried out after variation. Besides, to avoid the numerical ambiguity from the division of zero by zero, an odd number of mesh points in the integration over the gauge angles for particle number projection is adopted. More arguments have already been made in Ref.~\cite{Yao2009_PRC79-044312}.

%In the present investigations, we have not observed the spurious divergencies while checking the convergence of the projected energy as a function of the number of mesh points. This might be connected to the fact that we do not carry out a variation after projection. The investigation of correction from the finite spurious contribution is beyond the scope of the present work.

\section{Results and discussions}\label{sec:results}
%对称情况说明
In MDCRHB model and its extensions, thanks to the usage of an unified axial symmetric basis, we can alleviate the computational burden by simply disregarding the matrix elements that violate the specific symmetries.
We break a symmetry if and only if it is an essential element in describing certain processes we are focusing on.
%The MDCRHB model is thus a versatile tool enabling us to explore various sections of the multi-dimensional energy surfaces with moderate computational cost. 
Therefore, four types of spatial symmetries can be imposed and associated with some non-zero quadrupole-octupole deformations:
(a) axial-reflection symmetry with only $\beta_{20}$;
(b) non-axial but reflection symmetry with ($\beta_{20},\beta_{22}$); 
(c) axial symmetry but reflection asymmetry with ($\beta_{20},\beta_{30}$);
(d) non-axial and reflection asymmetry with ($\beta_{20},  \beta_{22}, \beta_{30},\beta_{32}$), %We use $K\pi, \slashed{K}\pi, K\slashed{\pi}, \slashed{K}\slashed{\pi}$
%to represent these cases, where $K(\pi)$ means Hamiltonian matrix elements with different $K(\pi)$ are separated and $\slashed{K}(\slashed{\pi})$ means they are mixed.
which are labeled as $K\pi, \slashed{K}\pi, K\slashed{\pi}$, and $\slashed{K}\slashed{\pi}$, respectively. 
%1DAMP

\begin{figure}[htbp]
	\centering
	\includegraphics[width=0.48\textwidth]{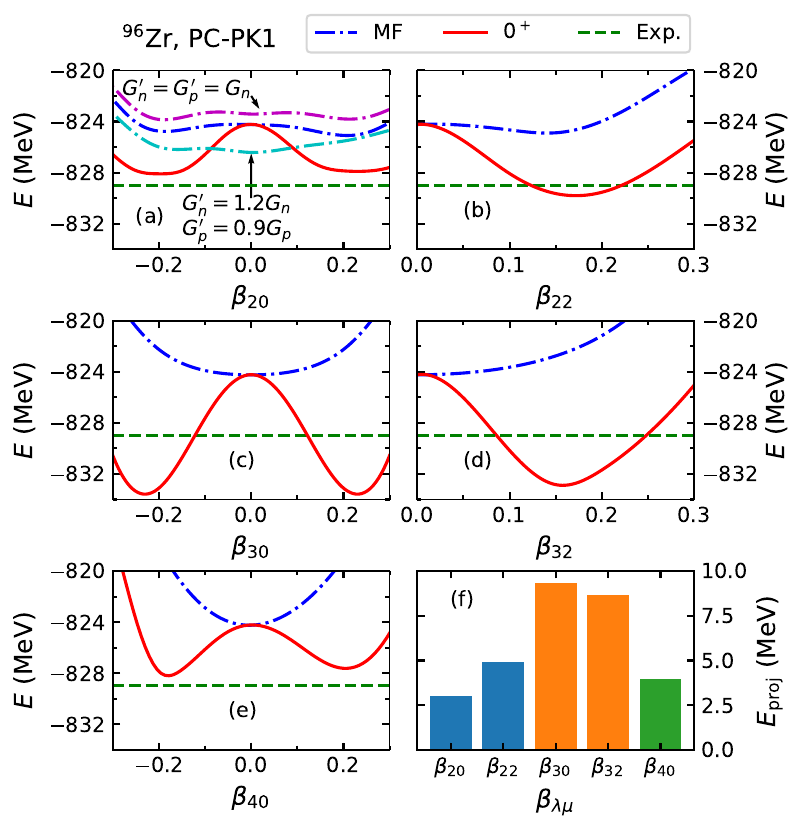}
	\caption{(Color online) The one-dimensional mean-field (dash-dotted lines) and projected $0^+$ (solid lines) potential energy surfaces of $^{96}$Zr with pure $\beta_{20}$ (a), $\beta_{22}$ (b), $\beta_{30}$ (c), $\beta_{32}$ (d) and $\beta_{40}$ (e) deformations, and projection energy as a function of $\beta_{\lambda\mu}$ (f).
If not specified on the figure, the pairing strengths are $G_p=815.36$ MeV$\cdot$fm$^3$ and $G_n=728.00$ MeV$\cdot$fm$^3$, respectively.
The experimental binding energy $E=-828.99$ MeV \cite{Wang2021_ChinPhysC45-030003} is shown as dashed line. 
}
	\label{fig:ZR96_PC-PK1_1DAMP}
\end{figure}

Fig.~\ref{fig:ZR96_PC-PK1_1DAMP}(a-e) shows the total energy of $^{96}$Zr as a function of $\beta_{20}$, $\beta_{22}$, $\beta_{30}$, $\beta_{32}$ and $\beta_{40}$, respectively, where all other deformation parameters are constrained to be zero.
The dash-dotted and solid lines correspond to the mean-field and projected $0^+$ PESs, respectively.
It is shown in Fig.~\ref{fig:ZR96_PC-PK1_1DAMP}(a) that the mean-field energy surface is very soft around the spherical shape. Any slight changes of the pairing strength may shift the location of the energy minimum. Specifically, with $G_n = 728.00$ MeV$\cdot$fm$^3$ and $G_p = 815.36$ MeV$\cdot$fm$^3$, we obtain two energy minima located at prolate ($\beta_{20}=0.21$) and oblate ($\beta_{20}=-0.19$) respectively with energy difference 0.32 MeV. If we enhance (quench) the neutron (proton) pairing strength by 20\%(10\%), the PES is still soft, but the energy minimum is shifted to the spherical shape. A proper description of the ground state needs to go beyond the mean-field approximation.

For $E(\beta_{22})$ shown in Fig.~\ref{fig:ZR96_PC-PK1_1DAMP}(b), the mean-field PES is very soft.
The projected $0^+$ energy minimum is located at $\beta_{22}=0.17$ and the corresponding binding energy is slightly overestimated.
The PESs $E(\beta_{30})$ and $E(\beta_{32})$ with octupole deformations are shown in Fig.~\ref{fig:ZR96_PC-PK1_1DAMP}(c) and (d). One can see that the energy minimum in the mean-field PES is shifted from spherical to octupole deformed state with $\beta_{30}=0.23$ or $\beta_{32}=0.16$ respectively after symmetry restorations. The energies gained from projection are 9.33 MeV and 8.65 MeV, respectively, which are consistent with the results of HFB model with Gogny interaction \cite{Tagami2015_JPG42-015106}.
For $E(\beta_{40})$ in Fig.~\ref{fig:ZR96_PC-PK1_1DAMP}(e), The energy minimum changes from spherical to hexadecapole shape after projections and we gain 3.9 MeV extra binding energy by symmetry restorations. 
We define the projection energy $E_{\rm proj}$ as
\begin{equation}
	E_{\rm proj}=E({\rm MF})_{\min}-E(0^+)_{\rm min}, 
\end{equation}
where $E(\rm MF)_{\rm min}$ and $E(0^+)_{\rm min}$ are the global energy minima on the mean-field and $0^+$ PESs under the same symmetry restriction respectively, and show them in Fig.~\ref{fig:ZR96_PC-PK1_1DAMP}(f). One can find $E_{\rm proj}$'s for octupole deformations are the dominant ones, which are larger than those for quadrupole and hexadecapole deformations.

\begin{figure*}
	\centering
	\includegraphics[width=1.0\textwidth]{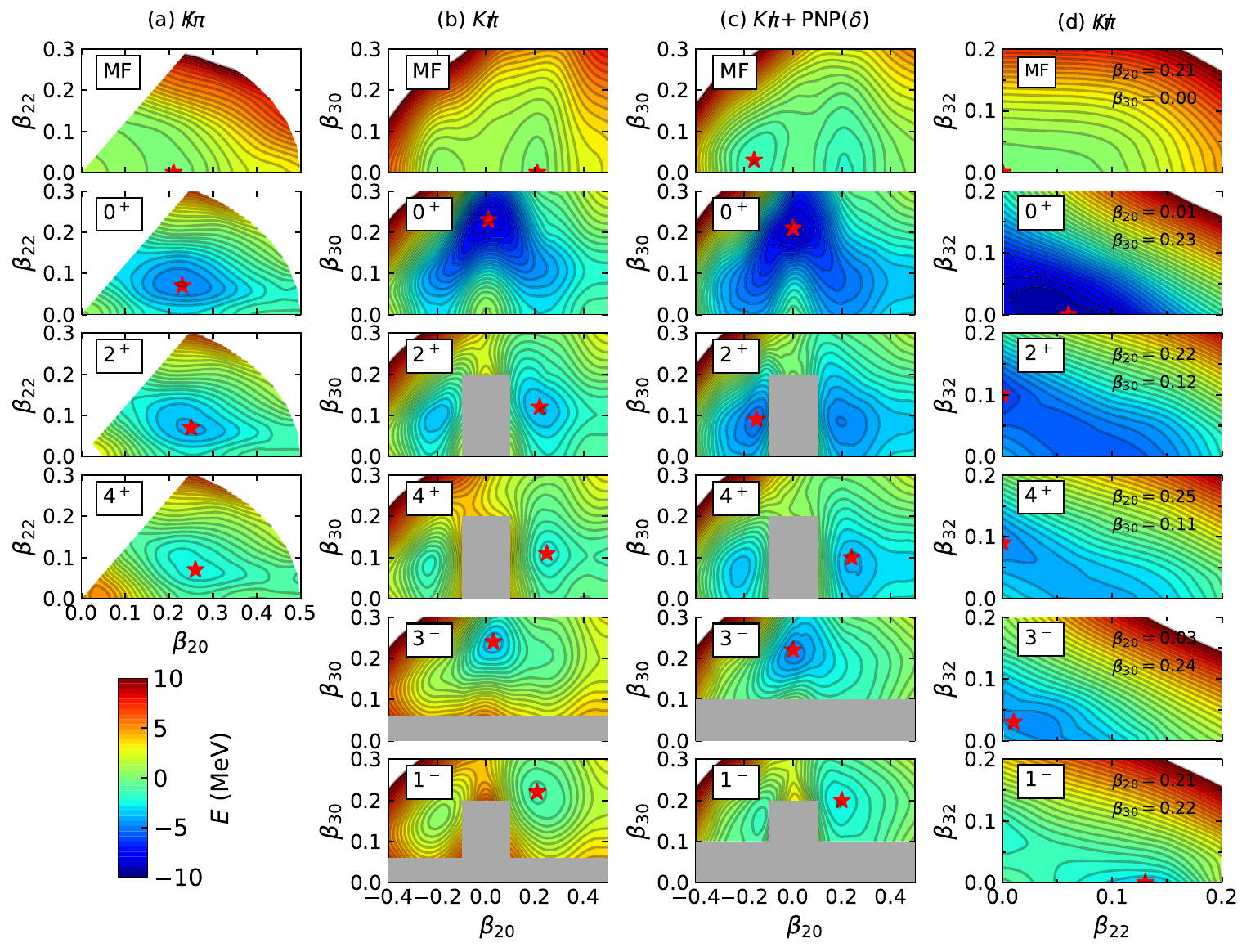}
	\caption{(Color online) The two-dimensional mean-field (MF), $J^\pi=0^+, 2^+, 4^+, 1^-$ and $3^-$ potential energy surfaces with (a) $\slashed{K}\pi$, (b) $K\slashed{\pi}$, and (d) $\slashed{K}\slashed{\pi}$ symmetry imposed.
The (c) $K\slashed{\pi}+$PNP($\delta$) case calculated by using three-dimensional harmonic oscillator expansion method and density-dependent $\delta$ pairing force developed in Ref.~\cite{Yao2015_PRC92-041304} are shown for comparison. 
The contours join points with the same energy, and the separation between adjacent contours is 0.5 MeV. On each PES the red star marks the corresponding global energy minimum. {The values marked in (d) are the constrained $(\beta_{20},\beta_{30})$ in this column, which are found in the global minima of (b).}
In each column the energies are relative to the mean-field ground state.
 }
\label{fig:ZR96_PC-PK1_2DAMP}
\end{figure*}

Next we turn to the two-dimensional PESs. 
The results are shown in Fig.~\ref{fig:ZR96_PC-PK1_2DAMP}. We also list the deformation $\beta_{\lambda\mu}$, projection energy $E_{\rm proj}$ and excitation energy $E_x$ corresponding to the global energy minima $E_{\rm  min}$ on each PES in Table \ref{tab:bet&Ex_from_2dPES}. We define $E_x$ as
\begin{equation}\label{eq:E_ex}
E_x(J^\pi)=E(J^\pi)_{\min}-E(0^+)_{\rm min},
\end{equation}
where we compare the energy minima on different PESs which usually take different shapes.
Note that this approximation only works for states with definite shapes.
For soft PESs we also need the GCM to take into account the shape fluctuations or shape coexistence.

%\begin{figure*}[htbp]

\begin{table*}[htbp]
	\setlength\tabcolsep{4.8pt}
	\centering
	\caption{ The deformations $\beta_{\lambda\mu}$ and excitation (projection) energies $E_x$ ($E_{\rm proj}$) of various $J^\pi$ (MF) states for $^{96}$Zr.
		$E(0^+)_{\rm min}$ denote the energy of the projected $0^+$ states.
		All excitation energies are relative to the $0^+$ states with the same symmetry.
		For $\slashed{K}\slashed{\pi}$ symmetry, $(\beta_{20}, \beta_{30})$ deformations are the same as those in the $K\slashed{\pi}$ global energy minimum.
		In the last line, the deformations inferred from the RHIC experiments~\cite{Zhang2022_PRL128-022301} are listed for comparison.
		The experimental excitation energies are taken from Ref.~\cite{Nudat2008}. All energies are in MeV. }
	\begin{threeparttable}
		\begin{tabular}{cccccccccccccc}
			\hline
			\hline
			Sym.
			&  \multicolumn{3}{c}{(a) {$\slashed{K}\pi$}}
			& \multicolumn{3}{c}{(b) $K\slashed{\pi}$}
			& \multicolumn{3}{c}{(c) $K\slashed{\pi}+$PNP($\delta)$}
			& \multicolumn{3}{c}{(d) $\slashed{K}\slashed{\pi}$}
			& ${\rm Exp.}$                                \\
			$E(0^+)_{\rm min}$ &  \multicolumn{3}{c}{$-829.978$}
			&  \multicolumn{3}{c}{$-833.616$}
			&  \multicolumn{3}{c}{$-832.329$}
			&  \multicolumn{3}{c}{$-834.546$}
			&  $-828.99$ \\
			\hline
			State & $\beta_{20}$  & $\beta_{22}$ & $E_{x/\rm proj}$ & $\beta_{20}$ & $\beta_{30}$ &  $E_{x/\rm proj}$  & $\beta_{20}$  & $\beta_{30}$ & $E_{x/\rm proj}$  & $\beta_{22}$  & $\beta_{32}$  &  $E_{x/\rm proj}$ & $E_x$ \cite{Nudat2008}\\
			\hline
			MF     & 0.21  & 0.00  & 4.89  % TA-RS
			& 0.21  & 0.00  & 8.53  % AS-RA
			&{$-0.16$} & {0.03}  & {6.94}
			& 0.00  & 0.00  & 9.46  % TA-TA
			
			& -\\
			$0^+$  & 0.23  & 0.07  & 0.00   % TA-RS
			& 0.01  & 0.23  & 0.00   % AS-RA
			& 0.00  & 0.21  & 0.00
			& 0.06  & 0.00  & 0.00  % TA-TA
			& 0.00 \\
			$2^+$  & 0.25  & 0.07  & 0.91  % TA-RS
			& 0.22  & 0.12  & 4.65  % AS-RA
			&$-0.15$& 0.09  & 3.41   %PNP
			& 0.00  & 0.10  & 3.06 % TA-TA
			& 1.75 \\
			$4^+$  & 0.26  & 0.07  & 2.31
			& 0.25  & 0.11  & 5.94
			& 0.24  & 0.10  & 4.70    % pnp
			& 0.00  & 0.09  & 4.59
			& 2.75\\
			$3^-$  & -     & -     & -   % TA-RS
			& 0.03  & 0.24  & 4.90 % AS-RA
			& 0.00  & 0.22  & 4.00  % PNP
			& 0.01  & 0.03  & 4.98 % TA-TA
			& 1.90\\
			$1^-$  & -     & -     & -
			& 0.21  & 0.22  & 7.09
			& 0.20  & 0.20  & 5.95  % pnp
			& 0.13  & 0.00  & 6.13  % ta-ra
			& 4.07 \\
			RHIC \cite{Zhang2022_PRL128-022301}
			&       &       &
			& 0.062 & 0.202 & \\
			\hline
			\hline
		\end{tabular}
		\label{tab:bet&Ex_from_2dPES}

		%\begin{tablenotes}
		%	\footnotesize
		%	\item {\color{red}The (c) $K\slashed{\pi}+$PNP($\delta$) case is calculated by using three-dimensional harmonic oscillator expansion method and density-dependent $\delta$ pairing force developed in Ref. \cite{Yao2015_PRC92-041304}.}
		%\end{tablenotes}
		
	\end{threeparttable}
\end{table*}

%E(b20,b22)
In $\slashed{K}\pi$ calculations, thanks to the six-fold symmetry with pure quadrupole deformations, 
we constrain the Hill-Wheeler coordinates $\beta_2=\sqrt{\beta_{20}^2+2\beta_{22}^2}, \gamma=\arctan(\sqrt{2}\beta_{22}/\beta_{20})$ with $\gamma\in [0,60^\circ]$ instead of $(\beta_{20},\beta_{22})$. Note that the octupole deformations will spoil the elegant six-fold symmetry and we constrain $\beta_{\lambda\mu}$'s directly in other cases. The mean-field PES is soft along the $\gamma$ direction and the two energy minima correspond to the prolate and oblate shape isomers in Fig.~\ref{fig:ZR96_PC-PK1_1DAMP}(a), respectively. After symmetry projections, there is only one energy minimum with $(\beta_{20},\beta_{22})=(0.23,0.07)$,  i.e.,
$(\beta_2,\gamma) = (0.25, 23^{\circ})$ in Hill-Wheeler coordinates. Thus the ground state
of $^{96}$Zr is triaxially deformed, consistent with
the Monte Carlo shell-model (MCSM) calculations~\cite{Kremer2016_PRL117-172503}.
The triaxial degree of freedom also appears in the energy minima of the $2^+$ and $4^+$ PESs.
These energy minima assume similar deformations, indicating that they belong to the same collective band.
The excitation energies $E_x(2^+)=0.91$ MeV and $E_x(4^+)=2.31$ MeV are lower than their corresponding experimental values $1.75$ MeV and $2.75$ MeV~\cite{Nudat2008}, respectively.

In Fig.~\ref{fig:ZR96_PC-PK1_2DAMP}(b) we present the mean-field and projected PESs $E(\beta_{20},\beta_{30})$ with the $K\slashed{\pi}$ symmetries.
On the mean-field PES, the two energy minima are located in $(\beta_{20},\beta_{30})=(0.21,0.00)$ and $(-0.19,0.00)$, respectively.
After symmetry projections, we find a single energy minimum at $(\beta_{20},\beta_{30})=(0.01,0.23)$ on the $0^+$ PES.
The symmetry restorations give a large energy correction $E_{\rm proj}=8.53$ MeV.
The projected $2^+$ and $4^+$ PESs have similar structures: they both exhibit energy minimum at $(\beta_{20}, \beta_{30})\approx(0.2, 0.1)$.
One may conjecture that these states together with another $0^+$ state with similar deformations belong to a typical rotational band.
To examine the existence of such $0^+$ state, we also need further GCM calculations.

On the $3^-$ PES in Fig.~\ref{fig:ZR96_PC-PK1_2DAMP}(b), we find a single energy minimum at $(\beta_{20},\beta_{30})=(0.03,0.24)$ with $E_x(3^-)=4.90$ MeV.
In contrary to the $2^+$ and $4^+$ states, its shape is close to that of the $0^+$ state.
We also present the $1^-$ PES.
%Then we plot the $1^-_1$ PES to determine whether the $3^-_1$ state belongs to a rotational band in which $1^-_1$ state is the band hand.
We find that the $1^-$ and $3^-$ minima have very different shapes.
The very small experimental $E_x(3^-)=1.90$ MeV and the deformations of the states point out the $3^-$ state may be a octupole vibrational collectivity~\cite{Iskra2019_PLB788-396} and the $0^+$ is the band head of octupole vibration instead of that of a nearly spherical band predicted by the MCSM calculations~\cite{Kremer2016_PRL117-172503}.

We also calculate $E(\beta_{20},\beta_{30})$ in $K\slashed{\pi}$ case with full particle-number projection (PNP) using the method developed in Ref.~\cite{Yao2009_PRC79-044312}. The results are shown in Fig.~\ref{fig:ZR96_PC-PK1_2DAMP}(c). Note that in these calculations we used different parameter settings compared to the other columns, such as the dimension of the basis space, the pairing force (in (c) we used a $\delta$-pairing force, while in the other columns we used a separable pairing force) and the particle-number corrections (in (c) we performed an exact particle-number projection, while in the other columns we used the approximate ansatz Eq. (\ref{eq:PNrestoration})).
For all $J^\pi$ states with the same symmetries imposed, the PESs are qualitatively similar. The only noticeable difference is that the approximate scheme in (b) gives prolate minima for mean-field and $2^+$ projected PESs, while the exact PNP method in (c) finds oblate minima on these PESs. The reason is that the PES of this nucleus is $\gamma$-soft and both axial-symmetric minima are nearly degenerate. A slight difference in the parameters can reverse the order of them. Thus, for more precise calculations the GCM must be included to mix the shapes.

In Fig.~\ref{fig:ZR96_PC-PK1_2DAMP}(d) we show the energy $E(\beta_{22},\beta_{32})$ of mean-field state and the states with projections onto different spin-parity as a function of $(\beta_{22},\beta_{32})$, while $(\beta_{20},\beta_{30})$ are fixed to the values corresponding to energy minima in Fig.~\ref{fig:ZR96_PC-PK1_2DAMP}(b). For mean-field PES, the energy minimum is located at $(\beta_{22}, \beta_{32}) =(0.0, 0.0)$, indicating that the non-axial deformations are not favored, consistent with the prediction of the RHB theory~\cite{Agbemava2016_PRC93-044304}, the HFB theories with Gogny interaction~\cite{Robledo2016_EPJA52-300} and Skyrme interactions~\cite{Cao2020_PRC102-024311}. 
In contrast, after projection onto spin-parity $J^\pi=0^+$, the energy minimum shifts to the deformations $(\beta_{20},\beta_{22},\beta_{30},\beta_{32}) = (0.01, 0.06, 0.23, 0.00)$.
However, the projected $0^+$ PES is soft against $(\beta_{22},\beta_{32})$ deformations. One may expect large shape fluctuations, the consideration of which is beyond the scope of this work. Nevertheless, our results show nontrivial contribution from both triaxial $\beta_{22}$ and tetrahedral $\beta_{32}$  deformations, which is consistent with the finding in Refs.~\cite{Kremer2016_PRL117-172503,Tagami2015_JPG42-015106} and the RHIC experiment~\cite{Jia2022_PRC105-044905}. Of course, this argument is based on the assumption that the location of the energy minimum in $(\beta_{20},\beta_{30})$ plane does not move when varying $(\beta_{22},\beta_{32})$. A more solid conclusion should be made based on the result of calculation for the states within the whole $(\beta_{20},\beta_{22},\beta_{30},\beta_{32})$ deformation space, which is too expensive to be carried out in this work.
Similarly, the above argument would also apply to all excited levels. 

%The excitation energies can be slightly changed when a full scan is performed.

\begin{figure}[htbp]
	\centering
	\includegraphics[width=0.46\textwidth]{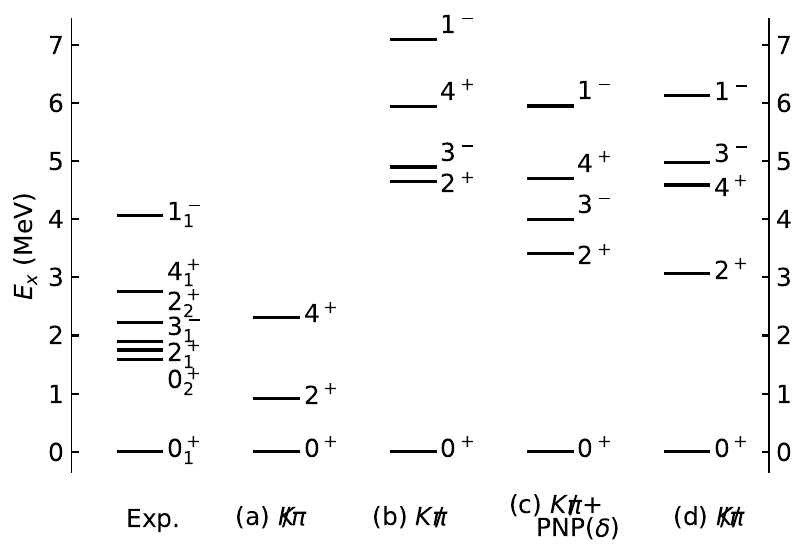}
	\caption{Comparison of the experimental low-spin $^{96}$Zr levels with the results of (a) $\slashed{K}\pi$, (b) $K\slashed{\pi}$ and (c) $K\slashed{\pi}+$PNP($\delta$) and (d) $\slashed{K}\slashed{\pi}$ calculations. In the $\slashed{K}\pi$ case, the levels belong to $\gamma$ band are not shown.
}
	\label{fig:ZR96_PC-PK1-levels}
\end{figure}

Next we examine the calculated energy levels and transition amplitudes. In Fig.~\ref{fig:ZR96_PC-PK1-levels}, the calculated low-lying excitation energies, defined in Eq.~(\ref{eq:E_ex}), are shown and compared with the experimental data~\cite{Nudat2008}.
The $E_x(2^+)$ with $\slashed{K}\pi$ symmetry is lower than the experimental value but the situation is opposite with $K\slashed{\pi}$ case. The excitation energies extracted from full PNP calculations are larger than the experimental values by about 2 MeV, lower than the case with particle-number restriction for about 1 MeV.
By breaking the axial symmetry, the energy levels $2^+$, $4^+$ and $1^-$ in $\slashed{K}\slashed{\pi}$ calculations are lowered compared with the results in the $K\slashed{\pi}$ calculations.
Conversely, the $3^-$ energy is almost unaffected by the triaxial deformations.
These level schemes are qualitatively similar to that from MCSM calculations~\cite{Iskra2019_PLB788-396}.

In the current framework, the electric multipole transitions are calculated between states that are projected out with the same deformations. In other words, the shift of the nuclear shapes with spin is neglected. Under this approximation, we calculate the $B(E3;3^- \rightarrow 0^+)$ using the symmetry-projected wave functions with intrinsic deformations $(\beta_{20},\beta_{30}) = (0.01, 0.23)$, corresponding to the energy minimum of $0^+$ state labelled with $K\slashed{\pi}$. The result is $20.68 \times 10^3~{\rm e}^2\,{\rm fm}^6$, which is comparable to the recent experimental value of $23(2) \times 10^3~{\rm e}^2\,{\rm fm}^6$~\cite{Iskra2019_PLB788-396}. Considering the fact that the energy minimum of the $3^-$ state coincidents with that of the $0^+$ state, as shown in Fig.~\ref{fig:ZR96_PC-PK1_2DAMP}(b), it is reasonable to compare this calculated value with the data. From the energy spectrum, the lowest $3^-$ state probably belongs to an octupole vibration band, the proper description of which requires the inclusion of the shape mixing. It is beyond the current framework. However, it is interesting to note that the $B(E3)$ is in surprisingly good agreement with experimental data.

We also calculate the mean-field and projected PESs with DD-PC1~\cite{Niksic2008_PRC78-034318} and PC-F1~\cite{Buervenich2002_PRC65-044308} parameter sets. The results are qualitatively the same as those we present here.

\begin{figure}[htbp]
	\centering
	\includegraphics[width=0.46\textwidth]{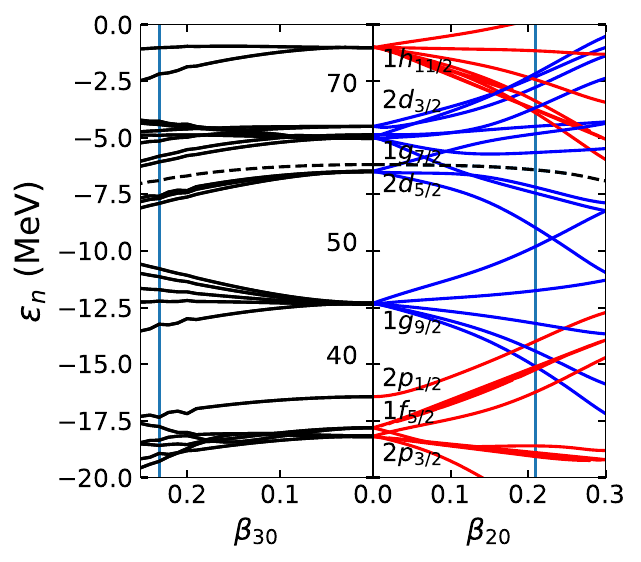}
	\caption{(Color online) Single-neutron levels of $^{96}$Zr as functions of pure $\beta_{20}$ (right half) and $\beta_{30}$ (left half) deformations.
The vertical lines on right and left halves correspond to global minima on the mean-field and $0^+$ PESs in Fig.~\ref{fig:ZR96_PC-PK1_2DAMP}(b), respectively.
The principal, orbital, and angular momentum quantum numbers $nlj$ for spherical symmetry are presented.
The dashed line denotes the Fermi level. }
	\label{fig:Zr96_PC-PK1_proton-SPL_b20-b30_Nf=10_Gp=1.12G0}
\end{figure}

Microscopically, the emergence of octupole collectivity stems from the specific shell structures of single-particle levels around
the Fermi level~\cite{Jahn1937_PRSA161-220,Butler1996_RMP68-349}.
In Fig.~\ref{fig:Zr96_PC-PK1_proton-SPL_b20-b30_Nf=10_Gp=1.12G0} we plot the mean-field single-neutron levels of $^{96}$Zr as functions of $\beta_{20}$ ($\beta_{30}$) deformation.
The single-proton levels below and around $Z=40$ are qualitatively similar and not shown here.
%四极关联:稳定的四极形变--d5/2 and g7/2
For the spherical shape $\beta_{20, 30}=0$, the neutron Fermi level lies between the $2d_{5/2}$ and $1g_{7/2}$ shells.
In the left half of Fig.~\ref{fig:Zr96_PC-PK1_proton-SPL_b20-b30_Nf=10_Gp=1.12G0}, we present
the dependence of the single-neutron levels on pure $\beta_{30}$ deformation. We see that the energy splittings caused by non-zero $\beta_{30}$ are small, and the spherical gaps at $N=40$, $50$, and $70$ are still remarkable.
In the right half of Fig.~\ref{fig:Zr96_PC-PK1_proton-SPL_b20-b30_Nf=10_Gp=1.12G0}, we show the splitting of the neutron energy levels due to a pure $\beta_{20}$ deformation.
Here the $1h_{11/2}$ state lying above the Fermi level splits into six levels.
Thus, neutrons have the possibilities to occupy the $1h_{11/2}$ orbital with considerable amplitude, leading to strong octupole coupling with the orbital $2d_{5/2}$.
As for single-proton energy levels, the $2p_{3/2}$ and $1f_{5/2}$ are the two states coupled via pure $\beta_{20}$ deformation while the $1g_{9/2}$ and $2p_{3/2}$ can be coupled via pure $\beta_{30}$ correlation.
The energy splitting for non-zero $\beta_{30}$ is again small compared with that for non-zero $\beta_{20}$, resulting in a large shell gap around the proton Fermi surface.
For non-zero $\beta_{20}$, some proton levels from the $1g_{9/2}$ and $2p_{3/2}$ states are close to each other near the Fermi surface.
Thus, the octupole deformation in $^{96}$Zr is dominated by the proton levels $2p_{3/2}\rightarrow 1g_{9/2}$ and neutron levels $2d_{5/2}\rightarrow 1h_{11/2}$.

\section{Summary}\label{sec:summary}
%方法+计算的物理量
The competitions of various shell structures near $A\simeq100$ create complex structures on the potential energy surfaces (PESs) of Zr isotopes.
In particular, it was found that both triaxial and octupole deformations play important roles in $^{96}$Zr, which presents a great challenge for calculations based on microscopic theories.
Recently we implement a unified framework called p-MDCRHB model to calculate the potential energy surfaces with various symmetries.
In this model both axial and reflection symmetries can be broken in the mean-field level and restored by projection-after-variation methods.
Various symmetries can be imposed to speed up the calculations.
With this model, we show that the  PESs of $^{96}$Zr have strong dependence on the angular momentum and parity.
Most mean-field and projected PESs are soft for triaxial deformation $\beta_{22}$ and tetrahedral deformation $\beta_{32}$.
We also extract the energy levels at the global minima on each projected PES and compare them with the experiments.
Our results suggest that both triaxial and octupole degrees of freedom are essential elements for precisely  describing the structure of the $^{96}$Zr nucleus.
A more solid conclusion cannot be made before the full symmetry-restored GCM calculation with the mixing of all types of deformed configurations is carried out. Work along this direction is in progress.

	\section*{Acnowledgements}
	
Helpful discussions with Dean Lee, Xiang-Xiang Sun, Kun Wang and Shan-Gui Zhou are
gratefully acknowledged. This work has been supported by the NSAF (Grant No. U1930403), the National Natural Science Foundation of China (Grants No. 12205057, No. 12275259 and No.12005082),  the promotion project of scientific research ability of young teachers in universities of Guangxi province (Grant No. 2022KY0055), the Central Government Guides
Local Scientific and Technological Development Fund Projects
(Grant No. Guike ZY22096024), the Jiangxi Provincial Natural Science Foundation (Grant No. 20202BAB211008). J.M.Y. is partially
supported by the Guangdong Basic and Applied Basic Research Foundation (Grant No. 2023A1515010936), and the National Natural Science Foundation of China (Grant No. 12141501). The results in this paper are obtained on the High-performance
Computing Cluster of ITP-CAS, the ScGrid of the Supercomputing Center, Computer Network Information Center
of Chinese Academy of Sciences, the Beijing Super Cloud Computing
Center (BSCC, http://www.blsc.cn/)
and  TianHe 3F. 
	
	%\begin{thebibliography}{00}
	%% \bibitem[Author(year)]{label}
	%% Text of bibliographic item
	%\bibliographystyle{elsarticle-num}
	%\bibliographystyle{elsarticle-harv}
	%\bibliographystyle{elsarticle-num-names}
	\bibliographystyle{elsarticle-num_20210104-sgzhou}
	%\bibliography{../../../information/refs/JabRef/sgzhou}
	%\bibliography{../../../Notes/bib/ref}
	\bibliography{./ref.bib}
	%\bibitem[ ()]{}
	
	%\end{thebibliography}
	%\end{CJK*}
\end{document}